\documentclass[twocolumn,3p,times,procedia]{elsarticle}

\usepackage{ecrc}
\usepackage{graphicx}
\usepackage{balance}
\usepackage{array,threeparttable}
\usepackage{url}

\volume{00}

\firstpage{1}

\runauth{}

\jid{procs}

\CopyrightLine{2022}{Published by Elsevier Ltd.}

\usepackage{amssymb}
\usepackage{amsmath}
\usepackage{multicol}
\usepackage{algorithm}
\usepackage{algpseudocode}
\usepackage{color}
\usepackage{enumitem,xcolor}

\usepackage[figuresright]{rotating}
\usepackage{bm}
\usepackage{caption}
\usepackage{titlesec}
\titleformat*{\section}{\small \bf}
\titleformat*{\subsection}{\small \em}
\titleformat*{\subsubsection}{\small \em}

\usepackage{fancyhdr}
\fancyheadoffset[RO,EL]{0pt}
\usepackage{geometry}
\begin{document}\small
\begin{frontmatter}




\dochead{}
\title{
\begin{flushleft}
{\LARGE Large language model empowered next-generation MIMO networks: fundamentals, challenges, and visions}
\end{flushleft}
}
 %

\author[]{ \leftline {Zhe Wang$^a$$^b$, Jiayi Zhang$^*$$^a$$^b$, Hongyang Du$^c$, Ruichen Zhang$^d$, Dusit Niyato$^d$, Bo Ai$^a$$^b$, Khaled~B.~Letaief$^e$}}

\address{ \leftline {$^a$State Key Laboratory of Advanced Rail Autonomous Operation, Beijing Jiaotong University, Beijing 100044, China}
\leftline {$^b$School of Electronics and Information Engineering, Beijing Jiaotong University, Beijing 100044, China}
  \leftline {$^c$Department of Electrical and Electronic Engineering, University of Hong Kong, Hong Kong SAR 999077, China}
  \leftline {$^d$College of Computing \& Data Science, Nanyang Technological University, 639798, Singapore}
  \leftline {$^e$Department of Electrical and Computer Engineering, Hong Kong University of Science and Technology, Hong Kong SAR 999077, China}
}

\cortext[]{Zhe Wang (email: zhewang\_77@bjtu.edu.cn), Jiayi Zhang (Corresponding author) (email: jiayizhang@bjtu.edu.cn), Hongyang Du (email: duhy@hku.hk), Ruichen Zhang (email: ruichen.zhang@ntu.edu.sg), Dusit Niyato (email: dniyato@ntu.edu.sg), Bo Ai (email: boai@bjtu.edu.cn), Khaled~B.~Letaief (email: eekhaled@ust.hk)}


\begin{abstract}
Next-generation Multiple-Input Multiple-Output (MIMO) is expected to be intelligent and scalable. In this paper, we study Large Language Model (LLM)-enabled next-generation MIMO networks. Firstly, we provide an overview of the development, fundamentals, and challenges of the next-generation MIMO. Then, we propose the concept of the generative AI agent, which is capable of generating tailored and specialized contents with the aid of LLM and Retrieval Augmented Generation (RAG). Next, we comprehensively discuss the features and advantages of the generative AI agent framework. More importantly, to tackle existing challenges of next-generation MIMO, we discuss generative AI agent-enabled next-generation MIMO networks from the perspective of performance analysis, signal processing, and resource allocation. Furthermore, we present two compelling case studies that demonstrate the effectiveness of leveraging the generative AI agent for performance analysis in complex configuration scenarios. These examples highlight how the integration of generative AI agents can significantly enhance the analysis and design of next-generation MIMO systems. Finally, we discuss important potential research future directions.

\end{abstract}


\end{frontmatter}


\section{Introduction}
\begin{figure*}[t]
\centering
\includegraphics[scale=0.23]{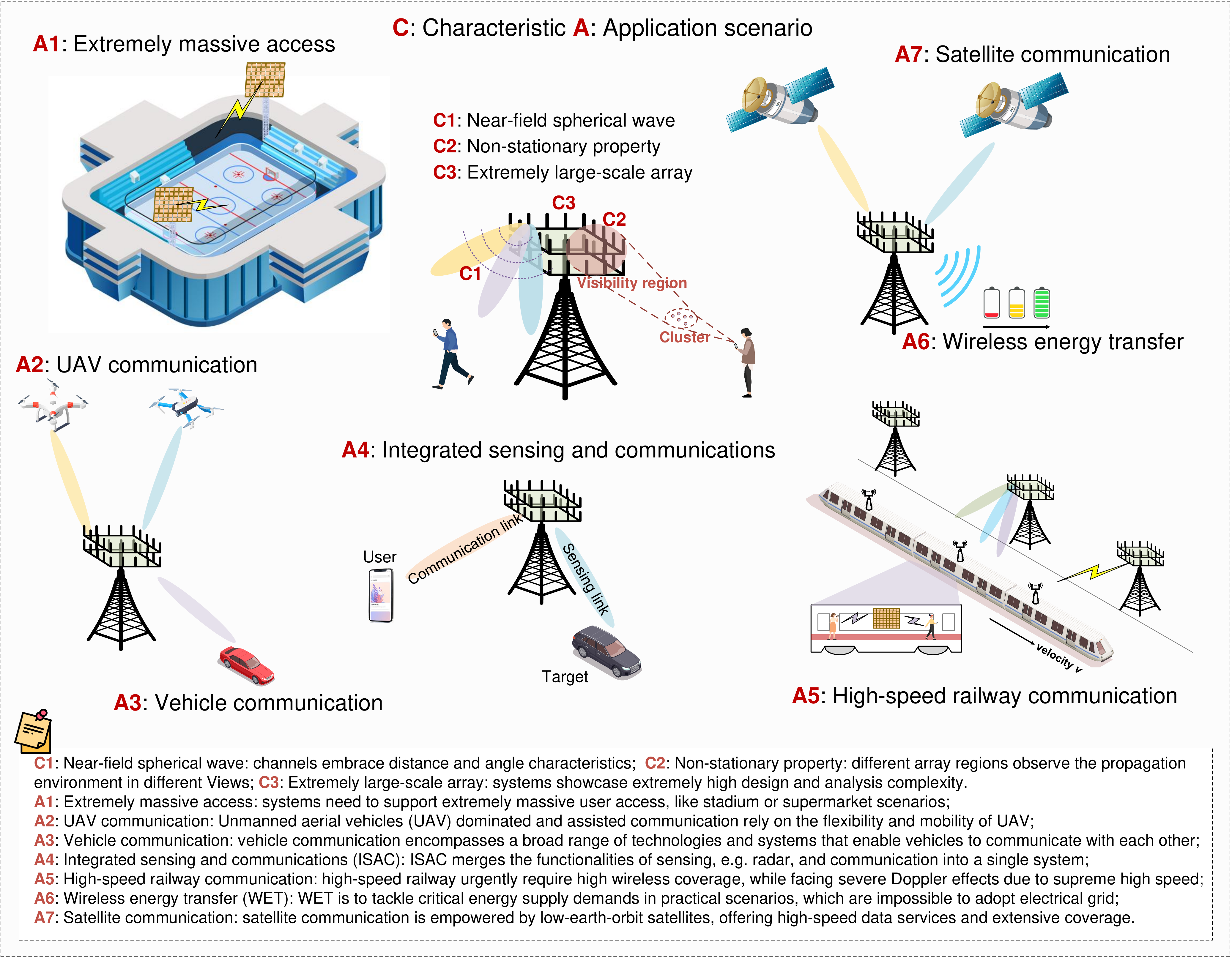}
\caption{Representative characteristics and application scenarios for next-generation MIMO. \label{MIMO}}
\end{figure*}

With the swift advancement of wireless communication, Multiple-Input Multiple-Output (MIMO) technology has experienced rapid growth to address the escalating demands of wireless communication and its extensive application scenarios \cite{9113273,10918608,11007274}. Serving as a cornerstone in the framework of wireless communication networks, the paradigms of MIMO technology have evolved significantly, diversifying from the conventional MIMO technology with a few antennas to Massive MIMO (mMIMO) technology with hundreds of antennas \cite{bjornson2024towards,OBETrans,chen2025channel}. mMIMO technology has been viewed as the promising empowered technology of the Fifth Generation (5G) of wireless communication networks. However, as future wireless communication demands continue to escalate \cite{8869705}, the next-generation MIMO technology is expected to be further studied and advanced. Extremely Large-Scale MIMO (XL-MIMO) technology, characterized by its deployment of thousands or even tens of thousands of antennas, is a representative next-generation MIMO technology, which has garnered significant research interest \cite{ZheSurvey,10143629}. In the following, the terminology `next-generation MIMO" refers to ``XL-MIMO" unless mentioned. The main characteristics and application scenarios for next-generation MIMO are illustrated in Fig. \ref{MIMO}. Compared with conventional mMIMO, the main differences for next-generation MIMO lie in another magnitude order of the number of antennas and near-field spherical wave characteristics.

However, the intricate system configurations and the numerous integrated communication scenarios pose significant challenges in the analysis and design of next-generation MIMO systems for both less-experienced researchers new to this particular topic and experienced researchers familiar with the technology. For less-experienced researchers, the complexity of formulating design problems and capturing the significant modeling characteristics can be overwhelming without extensive research experience. For experienced researchers, who focus on the sophisticated design and the facilitation of practical implementation, the encountered design problems are often as highly complex and intractable. This complexity not only necessitates a significant investment of time but also tends to result in a decrease in design efficiency or even errors. To tackle these formidable challenges, Artificial Intelligence (AI) technology can be applied in the next-generation MIMO system design relying on its intelligence and scalability \cite{bariah2024large,shafin2020artificial}. 

Generative AI (GAI) has attracted significant research interest and has been widely applied in practice due to its ability to generate customized content that meets specific user requirements \cite{2023arXiv230805384D,jovanovic2022generative,2024arXiv240216631Z}. A typical example of GAI in action is the development of large language models (LLMs), such as the generative pre-trained transformer (GPT) series by OpenAI \cite{shen2024large}. Despite their broad capabilities, LLMs may embrace limitations when they address highly specialized queries, constrained by the scope of their pre-training data \cite{chen2023benchmarking}. To tackle this challenge, the Retrieval Augmented Generation (RAG) approach has emerged as a promising solution, which can enhance the model's ability to provide specialized responses with the aid of targeted external datasets \cite{gao2023retrieval}. This combination between LLM and RAG technologies brings the formulation of the generative AI agent. The generative AI agent is a sophisticated technology designed to deliver adaptive and personalized content for specific fields, utilizing distinct, tailored external datasets \cite{zhang2024interactive}.

As discussed above, the analysis and design of next-generation MIMO systems face significant research challenges induced by the complicated system features and integrated scenarios. Notably,  empowered by the capabilities of the generative AI agent framework, the next-generation MIMO system analysis and design can be facilitated with enhanced efficiency and customization. Specifically, the generative AI agent can significantly provide comprehensive support in performance analysis, signal processing, and resource allocation. Specifically, with the aid of targeted external datasets, it facilitates a streamlined approach to formulating optimization design problems, offering innovative solutions to these problems, and conducting thorough evaluations of proposed design schemes based on specific requirements. These tailored steps encompass the intricate steps of identifying optimal parameters, navigating through complex design constraints, and achieving balanced trade-offs among various design objectives. Through leveraging the generative AI agent, designers and researchers can capture the complex landscape of next-generation MIMO system development with greater ease, ensuring that these advanced systems are both highly effective and versatile enough to accommodate the evolving needs of wireless communication technology.

In this paper, we investigate LLM-enabled next-generation MIMO system design. The main contributions are summarized as follows.
\begin{itemize}
\item We first comprehensively overview the development, fundamentals, and challenges of next-generation MIMO. Notably, the main challenges primarily arise from diversity of communication scenarios and complexity of system configurations, e.g., the number of antennas is extremely large leading to a large number of parameters/configurations to consider and optimize.
\item We investigate an emerging AI concept, a generative AI agent concept, which integrates the LLMs with the RAG to generate tailored and specialized contents. Then, we apply the generative AI agent framework to facilitate the next-generation MIMO analysis and design.
\item Two case studies are implemented to demonstrate the features and advantages of the generative AI agent in the next-generation MIMO analysis and design. Finally, potential future directions are discussed.
\end{itemize}

\section{Overview of Next-Generation MIMO}

XL-MIMO technology, which is the primary key technology toward next-generation MIMO, has been viewed as an important facilitator for 6G and the future generation of wireless communication networks \cite{ZheSurvey}. XL-MIMO is characterized by the deployment of an extremely large number of antennas, which could range from thousands to even ten thousand, in a remarkably compact space. In this section, we present basic fundamentals research challenges for the next-generation MIMO networks, from performance analysis, signal processing, and resource allocation.

\subsection{Performance Analysis}
A performance analysis framework is important to provide basics for further investigation and optimization design.

\subsubsection{Fundamentals}
With the further increase of the number of antennas, XL-MIMO systems must consider near-field spherical wave characteristics \cite{ZheSurvey,2024arXiv240105900L}, diverging from the far-field planar wave characteristic in conventional mMIMO. This shift implies that performance analysis frameworks developed for far-field mMIMO are inadequate for near-field XL-MIMO systems \cite{ZheSurvey}. A notable metric, the Effective Degrees of Freedom (EDoF), has attracted significant attention for its potential to approximate the number of dominant sub-channels, thereby offering a direct link to system capacity and intuitively demonstrating performance limits \cite{ZheSurvey}.

In XL-MIMO systems, proper evaluation metrics, such as the Bit Error Rate (BER) and Symbol Error Rate (SER), are vital to evaluate data transmission quality, directly impacting the reliability and efficiency of communication. The deployment of a vast antenna array can significantly improve signal quality, thereby reducing both BER and SER. Additionally, other key metrics include the Signal-to-Interference-plus-Noise Ratio (SINR), which assesses wireless communication link quality, and the Cram\'{e}r-Rao Bound (CRB), a lower bound on the variance of any unbiased estimate of a particular parameter. Indeed, it is vital to consider the proper performance metric and employ the corresponding analysis framework to characterize the system.

\subsubsection{Challenges} Some challenges should be pronounced both for researchers in the performance analysis of XL-MIMO.

\emph{(a). Analyses of Complex Scenarios}: 
To bring more valuable insights for the system performance, it is important to analyze practical but sometimes complex scenarios beyond simplistic model setups, such as non-parallel transceiver, multi-user, and diverse transceiver setups. Due to the more diverse system configurations for XL-MIMO systems, such as the more choices of antenna array, compared to the conventional MIMO systems \cite{ZheSurvey}, the analyses of complex scenarios for XL-MIMO would be much more complex than the conventional MIMO systems. For beginners with less research experience, it poses significant challenges to directly obtain insightful ideas for modeling these complex analysis scenarios due to diverse system configurations and numerous system parameters. Since it is necessary to have a profound understanding of the fundamental principles and techniques, which can facilitate the modeling of these intricate situations. Even for experienced researchers, although they have extensive knowledge and intuitive modeling capabilities in this research field, they may also have the risk of inadvertently overlooking minor modeling constraints. Such oversights can significantly impact the analysis and potentially lead to inaccurate or incomplete results.

\emph{(b). Various Integrated Scenarios}: 
The advancement of next-generation MIMO technology and evolving wireless communication contexts have ushered in emerging integrated scenarios like Integrated Sensing And Communication (ISAC), Space-Air-Ground Integrated Networks (SAGINs), and Wireless Energy Transfer (WET) \cite{2024arXiv240105900L}. Analyzing these scenarios demands properly selected performance metrics and a thorough understanding of their essential features for accurate framework development. Beginners may find it difficult to choose relevant metrics and grasp the essential aspects of these scenarios. For experienced researchers, it is time-consuming to capture key characteristics for studied scenarios, since these scenarios often exhibit numerous unique features and require a comprehensive analysis to identify and understand their underlying mechanisms.

\subsection{Signal Processing}
In next-generation MIMO systems, signal processing schemes are designed to enhance the system performance and meet various communication demands.
\subsubsection{Fundamentals}
In XL-MIMO systems, accommodating near-field spherical wave channel characteristics is pivotal for devising effective signal processing schemes, including channel estimation, beamforming, and beam codebook design. Near-field channels display polar sparsity, diverging from the angle sparsity in far-field channels based mMIMO, necessitating specialized channel estimation approaches \cite{ZheSurvey}. Furthermore, beamforming design mainly involves the receiving combining design and the transmitting precoding design. Receiving combining is implemented at the receiver to efficiently decode received signal based on different design goals, such as signal power maximization and interference minimization. Transmitting precoding is designed at the transmitter also with different goals to involve preprocessing the signal before transmission to enhance the network performance. Meanwhile, the beam codebook design is defined as the creation of a set of predefined beamforming vectors or beam patterns, which should efficiently balance the codebook size and system performance.

\subsubsection{Challenges}
To fully explore the potentials of next-generation MIMO systems and meet various wireless communication requirements, signal processing is important but facing many challenges for both new and experienced researchers: 

\emph{(a). Various Design Requirements}: 
Next-generation MIMO, especially XL-MIMO, integrates a wide spectrum of design parameters, scenarios, and objectives essential for high-efficiency and reliable communication. Taking the beamforming design as an example, there are numerous variables (e.g., digital and analog combining/precoding schemes), objectives (for instance, rate maximization, error minimization, and energy efficiency maximization), and scenarios (such as multi-user, ISAC, and UAV communications). These complex aspects pose significant challenges for beginners to formulate feasible optimization problems that accurately capture specific characteristics and meet diverse communication requirements. Furthermore, to enhance the practical implementation of studied designs, experienced researchers need to consider trade-offs between different objectives in various scenarios, such as the balance between the higher rate performance and increased energy consumption. Furthermore, the theoretical beamforming design should be practically implementable, requiring the consideration of real-world design requirements and constraints.

\emph{(b). Intractable Optimization Problems}: 
Signal processing design frequently involves complex optimization design problems with various objectives, variables, and constraints, rendering them intractable to solve. Many optimization methods with distinct properties and suitability for different scenarios are employed, including alternating optimization, Riemannian manifold optimization, genetic algorithm, and deep learning-aided algorithm. Therefore, first, it is challenging to choose an efficient optimization method suitable for their specific problem. Second, there is the intricacy of transforming non-convex optimization problems through various steps to render them solvable by optimization techniques. This conversion process, necessitated by the unique features of the optimization problems at hand, demands substantial time and effort.

\subsection{Resource Allocation}
In next-generation MIMO systems, due to the limited resources and massive connectivity requirement, the wireless communication resources need to be efficiently allocated.

\subsubsection{Fundamentals}
Resources in next-generation MIMO systems can be mainly categorized into power, pilot, and antenna resources, and all these resource types are critical to distinct problem domains such as power control, user initial access, and antenna selection. Power control schemes aim to allocate the system's limited transmission power efficiently to achieve goals like maximizing system capacity, enhancing minimum user performance, and improving energy efficiency. User initial access, the process of users first connecting to the network, should consider the constraints of pilot and antenna resources to minimize user interference or pilot contamination. As for the antenna selection, due to the extremely large array aperture, the antennas in XL-MIMO systems can be divided into a few sub-arrays. Based on this architecture, the antenna selection strategy should balance the trade-off between the system performance and processing complexity.

\subsubsection{Challenges}
When designing resource allocation strategies, some challenges need to be paid attention to.

\emph{(a). Tailored User Requirements}: 
It is important to notice that a valuable resource allocation strategy is often driven by practical user requirements. These requirements necessitate tailored resource allocation approaches.  For beginners, it is challenging to derive specific resource allocation goals and problems from practical user requirements, since it requires a deep understanding of both user requirements and system characteristics. Meanwhile, experienced researchers face the challenge of reconciling theoretically optimal resource allocation with practical constraints, striving for solutions that are both technically sound and practically implementable in real-world applications.

\emph{(b). Intricate Interrelationships with Other Designs}: 
Resource allocation strategies may have great effects on the further design and analysis of wireless communication systems. First, it is challenging to accurately formulate the system design optimization problem due to the complex relationship and constraint of the resource variables with other optimization variables. Second, the trade-offs between the different design objectives are sometimes inconsistent or even conflicting. However, due to the strong interrelationships between the resource allocation design and other designs, it is inevitable to jointly optimize these designs with a specific design goal.  

\textbf{$\bullet$ \emph{Lessons Learned}}: 
The analysis and design for next-generation MIMO systems encounter significant challenges due to the diverse communication scenarios and the complicated system configurations. These challenges require careful formulation and analysis of various design variables, goals, constraints, and scenarios in next-generation MIMO systems, which are challenging and time-consuming. Beginners may suffer from the lack of extensive research experience in this research field. Thus, it is challenging for them to accurately characterize fundamental system features when modelling the targeted systems and efficiently formulate feasible problems to effectively depict the key insights in next-generation MIMO systems. Furthermore, experienced researchers face challenges related to conducting in-depth studies and achieving practical scalability. These challenges can result in significant time consumption when efficiently formulating and solving highly sophisticated design problems.

\section{Generative AI Agent Framework for Next-Generation MIMO Networks}
With the rapid development of AI, GAI can be applied to facilitate and enhance the next-generation MIMO system design. In this section, we first overview generative AI agent and its collection of various techniques. We will then discuss generative AI agent-empowered next-generation MIMO system design.
\subsection{Overview of Generative AI Agent}

\subsubsection{Generative Artificial Intelligence}
GAI has been widely studied for its creative capability for generating new content, including texts, images, formulas, and designs. GAI embraces many promising concepts and technologies, such as the transformers, Generative Adversarial Networks (GANs), and Generative Diffusion Models (GDMs) \cite{2023arXiv230805384D}. Among them, transformers have revolutionized the machine processing of human language through their attention mechanism, enabling parallel input processing, exemplified by large language models such as ChatGPT. GANs operate based on the zero-sum game principle, which comprises a generator that produces synthetic data indistinguishable from real data, and a discriminator tasked with distinguishing between the synthetic and authentic data.

\begin{figure*}[t!]
  \centering
  \fontsize{8}{12}\selectfont
  \includegraphics[scale=0.6]{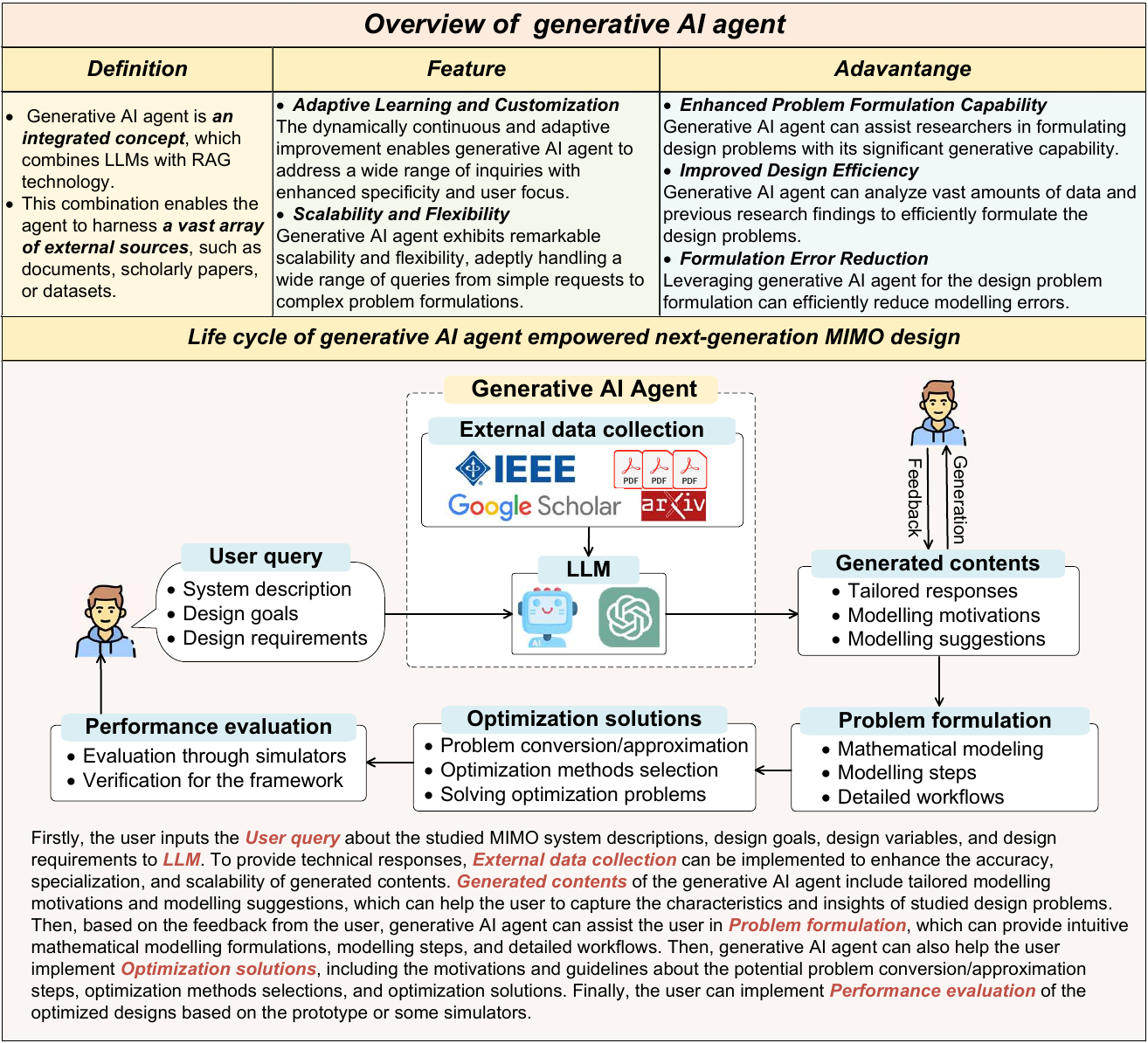}
\caption{Overview of generative AI agent: definition, feature, advantage of generative AI agent, and life cycle of generative AI agent empowered next-generation MIMO networks. \label{Table_Agent}}
   \vspace{0cm}
\end{figure*}

\subsubsection{Large Language Models}
LLMs, a key application GAI, are extensively applied in numerous fields. Built on transformer architectures with numerous parameters, LLMs are computational models designed to process, comprehend, generate, and interact with human languages. Trained on extensive textual data, LLMs can grasp and capture a wide range of linguistic nuances, patterns, and contexts. LLMs can efficiently address a variety of natural language processing tasks, including question answering, summarization, content refinement, and content generation, offering interactive and responsive solutions.

\subsubsection{Retrieval Augmented Generation}
LLMs' effectiveness is often constrained by their pre-training data, limiting their ability to address domain-specific and timely queries. To overcome this constraint, RAG has been introduced, which can enhance LLM responses by incorporating information from external knowledge bases. RAG represents a sophisticated advancement in natural language processing, combining LLMs' generative capabilities with external knowledge bases. This method improves the accuracy, relevance, and specialization of generated texts, increasing LLMs' applicability across various scenarios requiring current and specialized fundamentals.

\subsubsection{Features and Advantages of Generative AI Agent}
We now focus on the concept of the generative AI agent. Generative AI agent is a cutting-edge development of AI technology, which integrates LLMs with RAG technology with specialized and tailored generated content. This powerful combination enables the agent to harness a vast of external sources, such as documents, scholarly papers, or datasets. By processing this information, the generative AI agent is equipped to offer users optimized responses. This capability not only facilitates a deeper understanding of complex subjects but also enhances the process of problem-solving and system modeling, especially for next-generation MIMO, by providing personalized insights and actionable steps. The generative AI agent embraces the following main features.

\textbf{$\bullet$ \emph{Adaptive Learning and Customization}}:
The generative AI agent can generate detailed responses by analyzing queries from users, e.g., new or experienced researchers working on XL-MIMO technology, through internal and external databases, e.g., IEEE Xplore. It utilizes user interactions and feedback as a mechanism for continuous and adaptive improvement, which can enhance the response accuracy, relevance, and customization. Such dynamic adaptation enables the generative AI agent to become progressively skilled at offering accurate and insightful responses, addressing a wide range of inquiries with enhanced specificity and user focus.

\textbf{$\bullet$ \emph{Scalability and Flexibility}}:
The generative AI agent can adeptly handle a wide range of queries from simple requests to complex problem formulations. This versatile capability ensures the generative AI agent's scalability across diverse domains and user needs, underscoring its ability to adapt and respond to an extensive range of questions with precision and relevance.

Based on the definitions and features, the generative AI agent framework showcases the following advantages, especially in the next-generation MIMO system design:

\textbf{$\bullet$ \emph{Enhanced Problem Formulation Capability}}:
The generative AI agent can assist researchers in formulating design problems based on its significant generative capability with the aid of specified tailored external datasets. This capability is particularly beneficial for the next-generation MIMO networks, where complex considered scenarios and various optimized variables should be included. Thus, upon the promising generative capability, the generative AI agent can effectively boost the problem formulation capability.

\textbf{$\bullet$ \emph{Improved Design Efficiency}}:
The generative AI agent can analyze vast amounts of data and previous research findings to efficiently formulate the design problems in next-generation MIMO. This capability proves particularly advantageous for researchers, by enabling them to reduce the time taken for performance analysis and optimization formulation and significantly improve the next-generation MIMO design efficiency.

\textbf{$\bullet$ \emph{Formulation Error Reduction}}:
Leveraging the generative AI agent for the design problem formulation can efficiently reduce modeling errors. Given the intricate design requirements and variables requiring optimization in complex next-generation MIMO systems, there exists a considerable risk of overlooking essential modeling features, which may significantly affect the next-generation MIMO analysis and design. By referring to external and verified databases, the generative AI agent can validate modeling characteristics and optimization factors more comprehensively.  We summarize the basic fundamentals of the generative AI agent in Figure~\ref{Table_Agent}.

\begin{table*}[t!]
  \centering
  \fontsize{8}{12}\selectfont
  \caption{Overview of next-generation MIMO networks and generative AI agent assisted aspects.}
  \label{Comparisons}
  \includegraphics[scale=1.1]{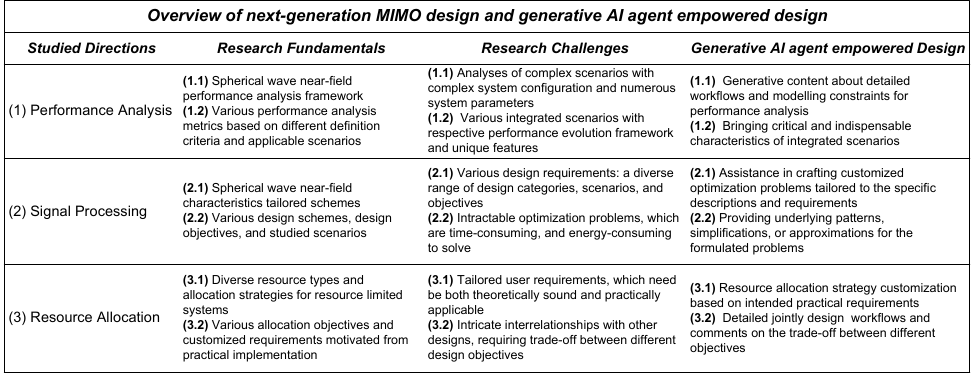}
   \vspace{0cm}
\end{table*}

\begin{figure*}[t]
\centering
\includegraphics[scale=0.7]{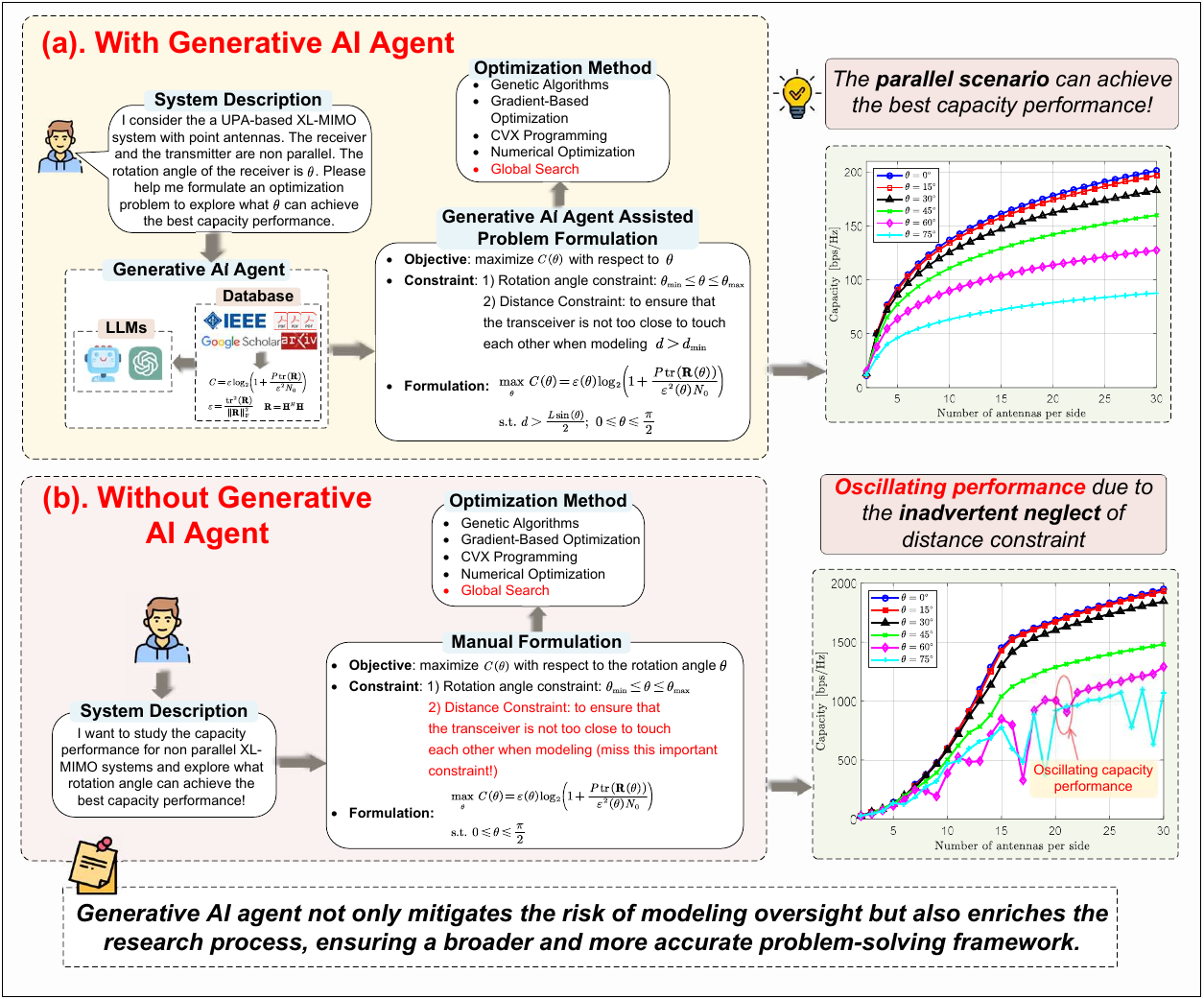}
\caption{Generative AI agent assisted capacity maximization for non-parallel UPA-based XL-MIMO systems. One square transmitting UPA surface and one square receiving UPA surface with similar physical sizes $10\lambda \times 10\lambda$ are considered, where the point antennas are uniformly distributed along the surfaces and $\lambda$ is the wavelength. The transmitting distances between the center point of the transmitter and the center point of the receiver for (a) and (b) are $30\lambda$ and $4\lambda$, respectively. And the Signal-to-Noise Ratio (SNR) $P/N_0=10 \ \mathrm{dB}$.\label{Rotation}}
\end{figure*}

\subsection{Generative AI Agent Empowered Next-Generation MIMO System Design}
We introduce the generative AI agent-empowered next-generation MIMO system design, the major steps of which are summarized in Figure~\ref{Table_Agent}.

\subsubsection{Performance Analysis}
The main challenges for the performance analysis are the complex analysis scenarios and various integrated scenarios. By analyzing MIMO system characteristics and requirements, the generative AI agent responds to users with detailed workflows and modeling constraints, which can improve the design efficiency and reduce the easily overlooked modeling errors.

For performance analysis of different scenarios, such as the next-generation MIMO-aided ISAC and WET, the generative AI agent can first assist the researchers to capture unique fundamental characteristics in particular aimed scenarios. Subsequently, it can recommend suitable performance metrics, thoughtfully chosen to showcase the distinct characteristics of the studied scenario while ensuring the analysis remains feasible and representative for next-generation MIMO systems. These approaches, facilitated by the generative AI agent, not only ensure that the distinctiveness of the integrated scenarios is adequately acknowledged and preserved but also guarantee that the performance analysis conducted is both comprehensive and tractable.

\subsubsection{Signal Processing}
The research challenges in signal processing are represented by a multitude of design requirements and the presence of complex and intractable optimization problems. The generative AI agent can first assist researchers in constructing customized design problems tailored to the specific MIMO configuration and studied scenario. This step can help researchers translate complex intended design requirements and various optimization challenges into intuitive and solvable problems. Then, the generative AI agent can further refine these problems into more manageable and solvable forms. This capability is achieved through a comprehensive analysis of the optimization design problem features, leveraging the generative AI agent's deep learning capabilities and extensive knowledge base.

\subsubsection{Resource Allocation}
The resource allocation mainly embraces the challenges of tailored user requirements and intricate interrelationships with other scheme designs. Accordingly, the generative AI agent is capable of customizing the resource allocation strategy design problem. This customization encompasses the formulation of optimization objectives, the careful selection of design variables, and the precise definition of design constraints, with particular emphasis on resource constraints like the limited bandwidth or pilot resources in next-generation MIMO systems. Furthermore, by thoroughly analyzing the complex interplay between resource-based variables and other design variables, such as the beamforming variable, the generative AI agent is equipped to offer detailed workflows to motivate the jointly optimized design for these variables. We summarize the fundamentals, challenges, and generative AI agent-empowered designs in Table~\ref{Comparisons}.

\begin{figure*}[t]
\centering
\includegraphics[scale=0.6]{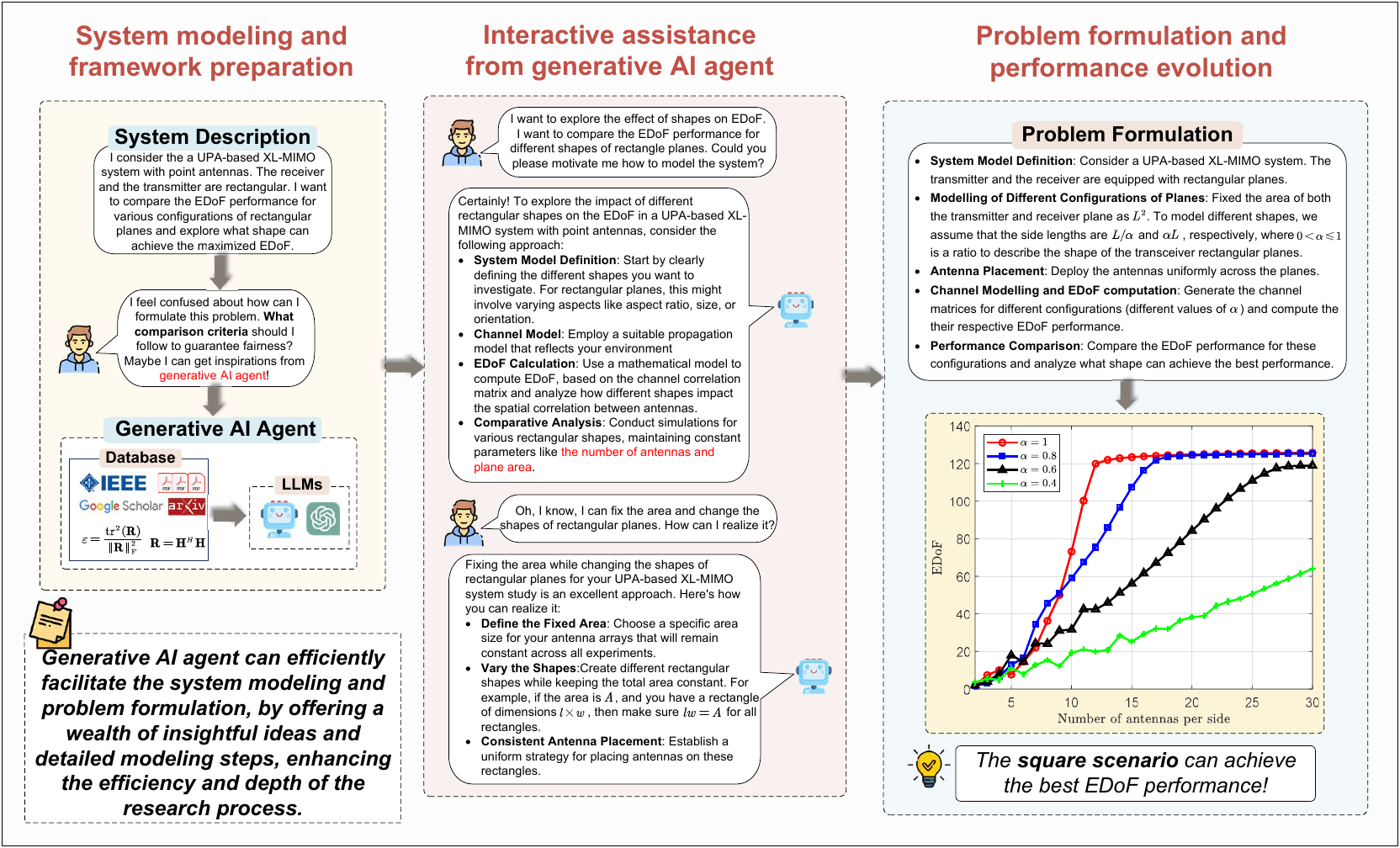}
\caption{Generative AI agent assisted EDoF maximization for rectangular UPA-based XL-MIMO systems with various shapes. We have $L=12\lambda$ and transmitting distance $d=10\lambda$. \label{Shape}}
\end{figure*}

\section{Case Studies}
In this section, we present use cases for generative AI agent-empowered XL-MIMO design. Specifically, we apply ChatGPT 4.0 as the agent\footnote{In this section, we use ``agent" to briefly represent ``generative AI agent".} to explore two cases of performance analysis\footnote{We provide detailed tutorials about case studies in this paper in https://zhewang77.github.io/GAIMIMO/.}.

\subsection{Capacity Maximization for Non-Parallel Transceiver}
Most studies on XL-MIMO systems consider an overidealistic parallel transceiver scenario \cite{[97]}. However, it is vital to explore the performance for the non-parallel transceiver scenario with a particular rotation angle $\theta$ considered, as varying  $\theta$ can significantly impact performance. We investigate how $\theta$ affects capacity in Uniform Plane Array (UPA)-based XL-MIMO systems, aiming to determine \textbf{\emph{when the system can achieve maximized capacity with different values $\theta$.}} This analysis is carried out based on the EDoF concept \cite{wang2024analytical}, which approximates dominant sub-channels' count and is closely linked to system capacity. The EDoF can be computed as $\varepsilon\triangleq{\mathrm{tr}^2\left( \mathbf{R} \right)}/{\left\| \mathbf{R} \right\| _{\mathrm{F}}^{2}} $, where $\mathbf{R}=\mathbf{H}^{H}\mathbf{H}$ is the channel correlation matrix with $\mathbf{H}$ being the channel matrix between the transceiver. Then, the capacity can be computed as $C=\mathrm{EDoF}\cdot \log _2( 1+\frac{\alpha P}{\mathrm{EDoF}^2N_0}) $, where $P$ is the transmitting power, $N_0$ is the noise power, and $\alpha$ is the overall channel gain \cite{wang2024analytical}.

Applying the generative AI agent framework and the extended database, i.e. a subset of papers from IEEE Xplore and arXiv, we formulate the capacity optimization problem for XL-MIMO systems considering rotation angles, as shown in Fig.~\ref{Rotation} (a). We extract the core generated contents by the agent, presented in the figures. The agent can extract important optimization constraints, such as the angle constraint and the transmitting distance constraint. Then, the agent is capable of aiding in the formulation of the capacity maximization problem. Emphasizing the transmitting distance constraint, it is vital for avoiding impractical scenarios and confounding results in the analysis. With the capacity maximization problem formulated, we then apply the global search method to explore how $\theta$ affects capacity performance. Simulations show that $\theta =0^{\circ}$ can achieve the maximized capacity, which means that the parallel transceiver configuration can achieve the best capacity performance.

Without the assistance from generative AI agent, as seen in Fig. \ref{Rotation} (b), formulating the optimization problem may face the risk of overlooking key constraints, such as the transmitting distance constraint. If the distance is smaller than $L\sin \left( \theta \right) /2$, the receiving plane would touch the transmitting plane, leading to an oscillating capacity phenomenon, which is impractical and misunderstanding. Notably, the assistance from the agent is helpful to comprehensively uncover minor modeling ideas and potential constraints. This help can facilitate a more thorough and accurate problem formulation. \textbf{\emph{Notably, generative AI agent not only mitigates the risk of modeling oversight but also enriches the research process, ensuring a broader and more accurate problem-solving framework.}} With the aid of the agent, in this case study, we study the effect of the non-parallel transceiver on system capacity performance and observe that the parallel transceiver scenario can achieve the maximized capacity performance.

\subsection{EDoF Maximization for Various Shapes of Transceiver}
We next study the effect of different rectangular shapes of the transceiver on the system performance and provide indispensable insights for practical implementation. Focusing on rectangular plane-based XL-MIMO, we investigate how different rectangular shapes affect EDoF performance, aiming to identify which shape yields the best EDoF performance. The variety of configurations and parameters, however, presents significant challenges for beginners to effectively understand the underlying motivations and formulate viable problems.

To tackle this challenge, the agent assists the researchers, especially the beginners, in modeling and formulating the optimization problem.
As shown in Fig. \ref{Shape}, by inputting system characteristics and descriptions of the aimed problems to the agent, the agent can adeptly provide researchers with insightful modeling ideas and comprehensive problem formulation steps. For instance, the agent advises to simulate various rectangular shapes with constant parameters including antenna number and plane area. With these descriptions, researchers can be motivated to explore different rectangular configurations by adjusting the shape ratio $0<\alpha \leqslant 1$, uniformly deploying antennas, and calculating EDoF performance for each shape. Simulation results indicate that a square shape ($\alpha=1$) achieves the maximized EDoF performance. This intuitive result can provide vital insights for the practical configuration implementation of next-generation MIMO systems. In this case, generative AI agent can effectively help sort out the key features and important workflows of the intended research problem. \textbf{\emph{As demonstrated above, generative AI agent can efficiently facilitate the system modeling and problem formulation, by offering a wealth of insightful ideas and detailed modeling steps, enhancing the efficiency and depth of the research process.}}

\section{Future directions}

\textbf{$\bullet$ Explainable AI (XAI)}:
Enhancing the explainability of LLMs is critical for the development of secure generative AI Agent-assisted wireless networks. This enhancement not only strengthens security and trustworthiness by delineating the AI's decision-making pathways but also significantly optimizes network functionalities. Specifically, by elucidating the LLMs' reasoning processes, operators can fine-tune resource allocation algorithms, ensuring that the most impactful factors are prioritized for network throughput and efficiency. For instance, in optimizing XL-MIMO configurations, a transparent understanding of how LLMs evaluate various antenna arrangements or signal processing techniques allows for more informed and effective decision-making. This level of detail in the models' operational logic enables targeted interventions, directly translating into enhanced network performance and reliability.

\textbf{$\bullet$ Generative AI Agent with Persistent Memory}:
In the future, the generative AI agent can be further designed with persistent memories. By leveraging persistent memories, the generative AI agent can adaptively learn and refine the understanding of researchers' research interests, research directions, and preferred methodologies over time. Indeed, the generative AI agent can effectively capture and retain valuable insights gleaned from previous interactions, enabling it to continuously evolve and tailor its specialized and efficient responses to facilitate the next-generation MIMO system design.

\textbf{$\bullet$ Digital Twins for Enhanced Validation}:
To bridge the divide between theoretical innovation and practical implementation in next-generation MIMO systems, a strategic pathway emerges by integrating the outputs of generative AI agents with digital twins. This approach facilitates the translation of theoretical advancements into real-world deployment, allowing for a more seamless and effective integration of cutting-edge MIMO technologies. Digital twins act as dynamic, virtual counterparts to physical systems, offering a robust platform for validating the performance of LLM and RAG-derived solutions across diverse scenarios. This method can facilitate extensive testing of MIMO configurations, providing valuable insights into their effectiveness, scalability, and real-world viability without relying on physical prototypes. 
Further advancing this approach, the ``autonomous lab'' concept utilizes robotics and cutting-edge technologies to automate XL-MIMO device deployment and testing. This shift towards automation minimizes human intervention, streamlining the validation process and enabling quicker, more efficient research and development iterations.

\section{Conclusions}
In this paper, we studied generative AI agent-enabled next-generation MIMO analysis and design. Firstly, the development, fundamentals, and challenges of the next-generation MIMO were reviewed. Then, we investigated generative AI agent and summarized its features and advantages. Next, generative AI agent-enabled next-generation MIMO analysis and design were motivated from performance analysis, signal processing, and resource allocation. To clearly showcase the advantages of generative AI agent, we provided two case studies to show how generative AI agent facilitates the next-generation MIMO analysis and design. Finally, some future directions were listed.


%


\bibliographystyle{elsarticle-num}
\balance
\bibliography{egbib}
~~~\\
~~~\\

\end{document}